# Dynamics of Soliton Cascades in Fiber Amplifiers

F. R. ARTEAGA-SIERRA,[1,*] A. ANTIKAINEN,[1] AND GOVIND P. AGRAWAL[1,2]

[1]*The Institute of Optics, University of Rochester, Rochester, New York 14627, USA*
[2]*Laboratory for Laser Energetics, 250 East River Rd, Rochester, NY 14623, USA*
*Corresponding author: f.arteaga-sierra@rochester.edu*

**We study numerically the formation of cascading solitons when femtosecond optical pulses are launched into a fiber amplifier with less energy than required to form a soliton of equal duration. As the pulse is amplified, cascaded fundamental solitons are created at different distances, without soliton fission, as each fundamental soliton moves outside the gain bandwidth through the Raman induced spectral shifts. As a result, each input pulse creates multiple, temporally separated, ultrashort pulses of different wavelengths at the amplifier output. The number of pulses depends not only on the total gain of the amplifier but also on the width of input pulses.**

Soliton propagation in erbium-doped fiber amplifiers (EDFAs) was studied in the early 1990s, mainly in the context of tele-communications [1–4]. Recent work has focused on super-continuum generation using a single or several cascaded amplifiers [5–8]. In this letter, we consider a single EDFA and study the evolution of femtosecond pulses launched with energy less than that required to form a fundamental soliton of equal duration. We show that multiple cascaded fundamental solitons are created at different distances within the amplifier. Each of them separates from the main pulse because of the Raman-induced frequency shift (RIFS) that moves them outside the gain bandwidth of the amplifier. Multiple fundamental solitons can also form in passive fibers, but their formation requires either modulation instability (long pulses) or soliton fission (short pulses), and both of these processes lead to solitons of different powers and durations [9]. The use of a fiber amplifier allows for the generation of multiple cascaded solitons of nearly the same widths and peak powers, without requiring modulation instability or soliton fission.

To study the evolution of short optical pulses inside an EDFA, we solve numerically the well-known generalized nonlinear Schrödinger equation [9, 10]. After adding a frequency dependent gain term this equation takes the following form in the frequency domain:

$$\frac{\partial \tilde{A}}{\partial z} - i[\beta(\omega) - \beta(\omega_0) - \beta_1(\omega - \omega_0)]\tilde{A} = \frac{g(\omega)}{2}\tilde{A}$$
$$+ i\gamma(\omega)\mathcal{F}\left(A\int_{-\infty}^{\infty} R(T')|A(z, T-T')|^2 dT'\right), \quad (1)$$

where $\mathcal{F}$ is the Fourier transform operator, $\tilde{A}(z,\omega) = \mathcal{F}(|A(z,t)|)$ is the Fourier transform of the complex pulse envelope $A(z,t)$, $\beta(\omega)$ is the propagation constant of the optical mode, $\beta_1 = \partial\beta/\partial\omega$ calculated at the carrier frequency w0 of the pulse, and $T = t - \beta_1 z$ is the time measured in a frame moving with the input pulse group velocity. The nonlinear effects are included through the parameter $\gamma(\omega)$ and a response function $R(t) = (1 - f_R)\delta(t) + f_R h_R(t)$ that includes the Kerr nonlinearity through the Dirac delta function and Raman nonlinearity through the commonly used form of the Raman response function $h_R(T)$ for silica with $f_R = 0.18$ [10]. The frequency dependent amplifier gain $G(z,\omega) = e^{g(\omega)z}$ is taken to be nearly flat over the amplifier bandwidth Ω and is included using a super-Gaussian profile:

$$G(\omega) = (G_0 - 1)\exp\left[-\left(\frac{(\omega-\omega_0)^4}{\Omega/2}\right)\right] + 1, \quad (2)$$

where $G_0$ is the maximum gain of the amplifier.

We solve Eq. (1) numerically with the split-step Fourier method [10] for a 20-m-long EDFA with its zero-dispersion wavelength ($ZDW$) at 1490 nm. Its gain spectrum is centered at 1550 nm ($\lambda_0 = 2\pi c/\omega_0 = 1550$ nm) and has a 40-nm gain bandwidth ($\Omega/(2\pi) = 5$ THz), common values for EDFAs used in telecommunications. Expanding $\beta(\omega)$ in a Taylor series around $\omega_0$, EDFA dispersion is included using $\beta_2 = -5.68$ ps$^2$/km and $\beta_3 = 0.13$ ps$^3$/km at 1550 nm. The nonlinear parameter has the form $\gamma(\omega) = \gamma_0 \omega/\omega_0$, where $\gamma_0 = \gamma(\omega_0) = 2$ W$^{-1}$/Km. The amplitude of the input pulse has the form $A(0,T) = \sqrt{P_0}\text{sech}(T/T_0)$ with $T_0 = 50$ fs (full width at half maximum about 88 fs). Its peak power $P_0$ is chosen such that the input soliton order is $N = T_0\sqrt{\gamma_0 P_0/|\beta_2|} = 0.7$, resulting in a peak power of 500 W. The peak gain $G_0$ of the EDFA is specified in units of dB/m and is varied from 0 to 4 dB/m. Figure 1 shows the temporal evolution of the 88-fs pulse over 20 m for $G_0 = 1, 2,$ and 4 dB/m. For $G_0 = 1$ dB/m, the central part of the input pulse forms a fundamental soliton ($N = 1$) after 3 m, and its spectrum begins to red-shift because of the RIFS [10], resulting in bending of the soliton trajectory owing to a reduction in its speed relative to the input pulse. The RIFS is relatively large for the soliton because its width is a fraction of the input pulse width. Amplification of this soliton stops after its spectrum moves out of the amplifier bandwidth, but the pulse remnants at the original location continue to be amplified as seen in Fig. 1(a). We even see the formation of a second soliton at a distance of about 15 m. Indeed, for a higher gain of $G_0 = 2$ dB/m in Fig. 1(b) we observe that multiple cascaded solitons form at different distances, and their trajectories bend toward the right because of the RIFS of each soliton. Each of these solitons also sheds some energy in the form of a dispersive wave (DW), as seen in Fig. 1(b). The situation becomes much more complex in Fig. 1(c) where the amplifier gain is increased to 4 dB/m. A large number of cascaded solitons emerge, together with their DWs that travel at different speeds and occasionally collide with the solitons. Even collisions of two neighboring solitons occur for an EDFA with such a high gain. One may wonder how the pulse spectrum evolves inside a fiber amplifier. Figure 2 shows the spectral and temporal evolutions for an amplifier with a gain of 3 dB/m. We clearly see the red-shifted spectral bands of the first two solitons formed at distances of about 2 and 6 m. Beyond that several solitons emerge in rapid succession so that their spectra overlap. At the amplifier output, a kind of supercontinuum is formed with two broad bands on each side of the input spectrum. The band on the red side belongs to multiple fundamental solitons, and the one on the blue side to the corresponding DWs. It is noteworthy that, unlike in conventional supercontinuum generation, the spectral broadening here is not based on soliton fission because the pulse energy never reaches the level that can support even a second-order soliton. Instead, the spectral broadening is solely due to RIFS, DW generation, and interactions among DWs and solitons.

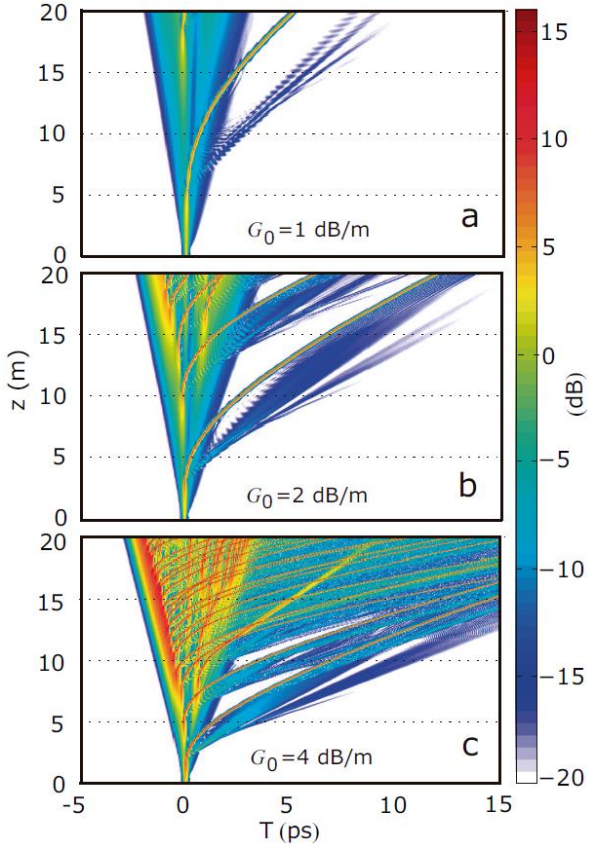

**Fig. 1.** Temporal evolution of of a 88-fs pulse over 20 m of active fiber for (a) $G_0 = 1$, (b) 2, and (c) 4 dB/m. Bent trajectories show cascaded red-shifted solitons forming at different locations inside the fiber amplifier.

It is clear from Figs. 1 and 2 that a single optical pulse propagating inside an optical amplifier can produce a cascade of ultrashort fundamental solitons, whose wavelengths are different at the amplifier output because each soliton forms at a different distance before experiencing the RIFS through intrapulse Raman scattering.

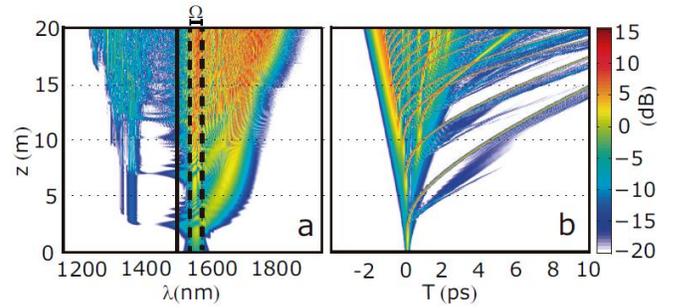

**Fig. 2.** Spectral (a) and temporal (b) evolution under the conditions of Fig. 1 with $G_0 = 3$ dB/m. The ZDW of the fiber is marked by a black line, and the dashed lines show the gain band.

The dynamics of these solitons exhibit rich behavior because of the simultaneous presence of a DW associated with each soliton and the possibility of collisions between two solitons or between a soliton and a DW. The number of solitons created can be controlled by varying the rate of amplification and the length over which pulse amplification occurs.

As an example, we study the case in which the optical gain exists only over the first few meters, i.e., an active section is followed by a passive fiber section. Figure 3 shows the temporal and spectral evolutions in two cases with $G_0 = 3$ dB/m. The gain is turned off after 2.5 m in the top row but after 7 m in the bottom row. As seen in the figure, only a single soliton forms in the low-gain case. The soliton decelerates even in the passive section and moves slower compared to the pulse remnants because of the RIFS [11]. A DW also appears because of energy transfer from the soliton at a phase-matched frequency during the process of spectral shifting [see parts (a) and (c)]. Since both the DW and the soliton spectra are distinct from the spectrum of input pulse, we can approximately calculate the soliton order by fitting a hyperbolic secant to the pulse remnants. The results are shown in Fig. 3(b), where we see that N exceeds 1 in the active section but drops to well below 1 in the passive section. In contrast, if the gain is kept on for a longer distance (parts d to f), the remnants continue to be amplified causing a second fundamental soliton to be generated at about 7.5 m, as seen in Figs. 3(d) and 3(f). The corresponding soliton order of the pulse remnants is shown in Fig. 3(e).

It should be clear by now that the number of fundamental solitons at the amplifier output depends heavily on the total gain $G_0$ of the amplifier. Figure 4 shows the number of solitons formed at the end of a 20-meter-long fiber amplifier (no passive section) as a function of G0 for two different widths of the input pulse. The solitons were counted manually from the evolution traces such as the ones shown in Fig. 1. In order for a pulse to be counted as a soliton, we required that it had started to separate itself from its surroundings and that its spectrum had begun to red-shift. Hence, we interpret Fig. 1(a) as showing only one soliton, as the formation of the other soliton amidst the pump remnants is not quite complete yet. Figure 1(b) shows six solitons, all of which are clearly red-shifting and moving slower with respect to the background.

As expected, Figure 4 shows that the number of solitons at the amplifier output increases with the total gain. For low gains ($G_0 < 22$ dB), a low-energy input pulse forms a soliton within few meters as N approaches 1 and then retains its soliton nature by reshaping itself to become shorter to account for the lack of initial energy. This phenomenon of adiabatic soliton compression is well known [1, 4]. It should be stressed that energy for the soliton comes mostly from the central region of the input pulse. After a certain gain threshold that pends on the input pulse duration, the gain is large enough to amplify and reshape the pulse remnants (mostly pulse wings) into another soliton, after the the first soliton has moved away because of its slowing down through the RIFS. This process repeats as $G_0$ is increased, and even more solitons are formed. The leading portion of the input pulse accounts for the formation of most solitons (see Figs. 1 and 2). This is because the RIFS-induced deceleration causes each soliton to lag behind and overlap with the trailing portion. As the solitons slow down, they deplete the trailing edge through nonlinear interactions. As a result, remaining pulse energy becomes heavily concentrated near the leading edge. Figure 4 shows that the number of solitons $n_s$ at the amplifier output also depends on the width of input pulses. For short pulses ($T_0 = 50$ fs), $n_s$ increases almost linearly with $\ln(G_0)$, or exponentially with $G_0$. This dependence becomes superlinear (or super-exponential) for wider pulses with $T_0 = 500$ fs. One can understand this feature as follows. For wider pulses the energy is more spread in time. Therefore, it takes a longer distance for a soliton formed near the leading edge of the pulse to reach its trailing edge. If the gain is large enough, the trailing edge of the input pulse can have enough time to create a soliton before it is consumed by the decelerating soliton. By the time the soliton from the leading edge reaches the newly formed trailing-edge soliton, their frequency separation is too large for the two solitons to collide and interact nonlinearly. As a result, both solitons survive and separate from the input pulse. The solitons formed in the trailing region of the long input pulse are responsible for the superlinear behavior seen in Fig. 4.

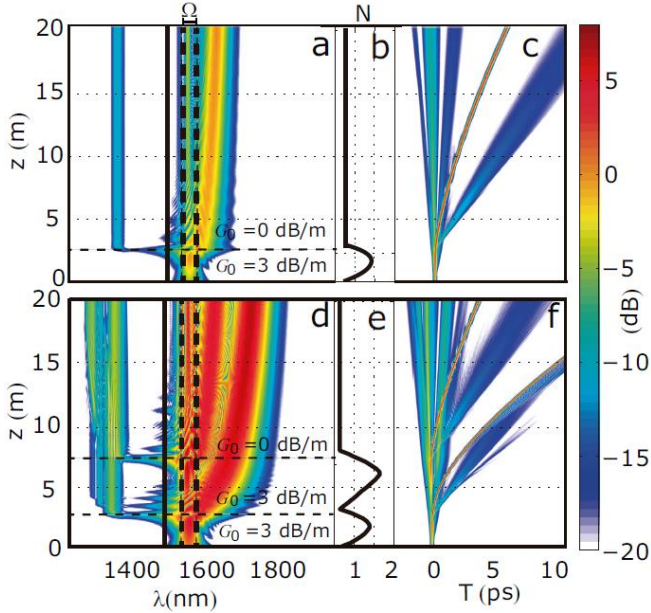

**Fig. 3.** Spectral [(a) and (d)] and temporal [(c) and (f)] evolution when the fiber is active over 2.5 m (top row) and 7.5 m (bottom row). The ZDW of the fiber is marked by a black line, and the vertical dashed lines show the gain band. The soliton order N of the pulse remnants is shown in parts (b) and (e).

The total gain also affects the spectral extent of the output. To study the effect of gain on the output spectrum, we define the spectral range $S_r$ as the difference between the largest and smallest frequencies for which the spectral power is below 50 dB of the maximum value. Note that this definition allows for gaps in the spectrum and should not be thought as the bandwidth of the output spectrum. Rather, $S_r$ is the total spectral range covering both the blue-shifted DWs and the red-shifted solitons. Figure 5 shows how $S_r$ at different distances of the active fiber

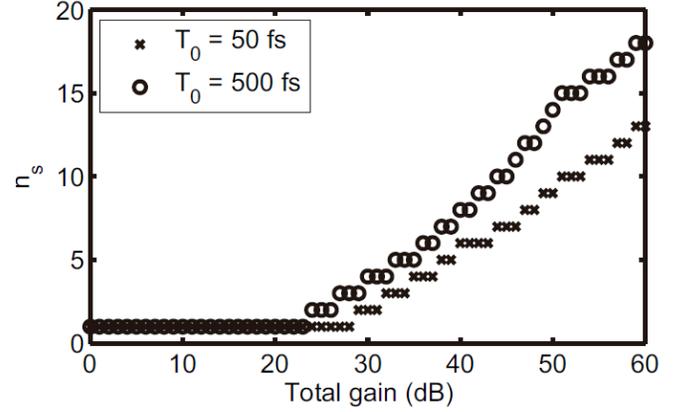

**Fig. 4.** Number of fundamental solitons ($n_s$) at the output of a 20-m-long amplifier as a function of total gain for two input pulses of different widths launched with N = 0.7.

depends on the total gain $G_0$ for the same two different pulse durations used in Fig. 4.

One can identify several different regions in Fig. 4. The spectral range is below 20 THz in the blue region where the pulse evolves to form a fundamental soliton that slowly red-shifts through the RIFS. The transition to the green region indicates the emission of a blue-shifted DW that increases $S_r$ to nearly 50 THz (or 400 nm). When the gain is sufficient, the first forming soliton continues to be amplified before it leaves the gain window all the way up to the point where it needs to readjust by shedding off a DW. Further propagation gradually extends $S_r$ because the RIFS increases the frequency separation between the DW and the soliton. Indeed, $S_r$ is close to 70 THz in the yellow region in Fig. 4.

For the short-pulse case shown in part (a) a second abrupt change occurs when the gain is high enough (red region for $G_0 > 40$ dB). The physical reason for this change is related to reflection of a DW from the moving index boundary created by a decelerating soliton formed later. As is well known, such temporal refections cause the DW to blue-shift further [12, 13] and eventually extend the spectral range to beyond 100 THz. The same physical processes occur in the case of longer pulses shown in Fig. 4(b) with some differences. First, the increase in the spectral extent is more gradual after the first DW generation. It can be attributed to interactions between two solitons and between a soliton and a DW. Second, extension to the red side can also happen through the formation of an abnormally redshifted soliton (example of an optical rogue wave) because of in-phase collisions of solitons [14–16].

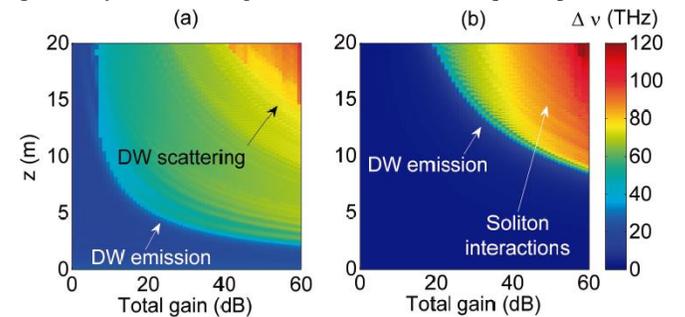

**Fig. 5.** Spectral range (color coded) as a function of propagation distance and total gain for input pulses with (a) $T_0 = 50$ and (b) 500 fs (N = 0.7 in both cases). Different nonlinear processes responsible for spectral changes are indicated.

In conclusion, we investigated numerically the propagation of short optical pulses (width < 1 ps) inside fiber amplifiers launched with less energy than required to form a soliton of equal duration. It was shown that the amplification leads to a cascade of independent fundamental solitons that appear at the amplifier output as temporally separated solitons of different wavelengths. The cascading process has its origin in the RIFS that red-shifts the spectrum of solitons while also slowing them down. The associated spectral broadening was attributed to soliton interactions and DW generation. The leading portion of the input pulse was shown to be responsible for the generation of vast majority of solitons for ultrashort pulses but the trailing part was also found to generate solitons for wider input pulses. We also found that the number of solitons at the fiber output depends not only on the total gain but also on the width of the input pulse.

Even though we focused on an EDFA with a 40-nm gain bandwidth in this work, our results are more general and should apply to all fiber amplifiers, as long as the dispersion is anomalous within the gain bandwidth. Our results are interesting from a fundamental perspective but they also point to a potential application. Temporally separated pulses of different wavelengths are often required in practice. Our results show that a fiber amplifier can be used to produce such pulses. Moreover, relative delays and wavelengths of different pulses are controllable through the length and gain of the amplifier and the width and peak power of input pulses.

**REFERENCES**

1. M. Nakazawa, K. Kurokawa, H. Kubota, K. Suzuki, and Y. Kimura, Appl. Phys. Lett. **57**, 653 (1990).
2. M. Nakazawa, Y. Kimura and K. Suzuki, IEEE J. Quantum Electron., **26**(12), 2103 (1990).
3. G. P. Agrawal, IEEE Photon. Technol. Lett., **2**(12), 875 (1990).
4. G. P. Agrawal, Phys. Rev. A **44**, 7493 (1991).
5. P. H. Pioger, V. Couderc, P. Leproux and P. A. Champert, Opt. Express **15**, 11358 (2007).
6. R. Song, J. Hou, S. Chen, W. Yang, and Q. Lu, Opt. Lett. **37**, 1529 (2012).
7. J. Swiderski, M. Maciejewska, Appl. Phys. B **109**, 177 (2012).
8. M. Tao, T Yu, Z. Wang, H. Chen, Y. Shen, G. Feng and X. Ye, Sci. Rep. **6**, 23759 (2016).
9. J. M. Dudley, G. Genty, and S. Coen, Rev. Mod. Phys. **78**, 1135 (2006).
10. G. P. Agrawal, Nonlinear Fiber Optics, 5th ed. (Academic Press, 2013).
11. N. Akhmediev and A. Ankiewicz, Solitons: Non-linear pulses and beams (Chapman & Hall, 1997).
12. A. V. Gorbach and D. V. Skryabin, Phys. Rev. A **76**, 053803 (2007).
13. B. W. Plansinis, W. R. Donaldson and G. P. Agrawal, Phys. Rev. Lett. **115**, 183901 (2015).
14. A. Antikainen, M. Erkintalo, J. M. Dudley, and Goery Genty, Nonlinearity **25**(7), R73 (2012).
15. R. Driben and I. Babushkin, Opt. Lett. **37**(24), 5157 (2012).
16. J. M. Dudley, F. Dias, M. Erkintalo, and G. Genty, Nature Photon. **8**, 755 (2014).